\documentclass[12pt,epsf,epsfig,psfig]{article}
\usepackage{graphicx}
\usepackage{epsfig}
\usepackage{cite}
\def\C{c{\bar c}}
\def\B{b{\bar b}}

\def\Q{Q{\bar Q}}

\def\NP{{ Nucl.\ Phys.\ }}
\def\PL{{ Phys.\ Lett.\ }}
\def\PR{{ Phys.\ Rev.\ }}

\def\ZP{{ Z.\ Phys.\ }}
\def\EPJ{{Eur.\ Phys.\ J.\ }}
\def\be{\begin{equation}}
\def\ee{\end{equation}}
\def\lsim{\raise0.3ex\hbox{$<$\kern-0.75em\raise-1.1ex\hbox{$\sim$}}}
\def\gsim{\raise0.3ex\hbox{$>$\kern-0.75em\raise-1.1ex\hbox{$\sim$}}}

\begin{document}

\parindent=0pt 

January 2015 \hfill BI-TP 2015/04

\vskip1cm

\centerline{{\Large \bf Quarkonium Binding and Entropic Force}}

\vskip1cm

\centerline{\large \bf Helmut Satz}

\bigskip

\centerline{Fakult\"at f\"ur Physik, Universit\"at Bielefeld}

\centerline{D-33501 Bielefeld, Germany}

\vskip1cm

\centerline{\large \bf Abstract:}

\bigskip

A $\Q$ bound state represents a balance between repulsive kinetic and
attractive potential energy. In a hot quark-gluon plasma, the interaction 
potential experiences medium effects. Color screening modifies the 
attractive binding force between the quarks, while the increase of entropy
with $\Q$ separation gives rise to a growing repulsion.
We study the role of these phenomena for in-medium $\Q$ binding and 
dissociation. It is found that the relevant potential for $\Q$ binding 
is the free energy $F$; with increasing $\Q$ separation, further binding 
through the internal energy $U$ is compensated by repulsive entropic effects. 

\vskip1cm

{\large \section{Introduction}}

The concept of entropic forces, emerging as a result of collective many-body
phenomena, has in recent times attracted increasing interest; see e.\ g.\ 
\cite{r1a,r1b,r1c,r2}. The effect of such forces arises from the thermodynamic 
drive of a many-body system to increase its entropy, rather than from a 
specific underlying microscopic force. This can also provide a way of 
studying the role of entropy maximization for a specific dynamic system 
immersed in a thermal medium. We here want to use this approach to address 
quarkonium binding and dissociation in a hot deconfined quark-gluon 
plasma \cite{r3}.

\medskip

The simplest approach to study $\Q$ binding in a medium of temperature 
$T$ is in terms of the Schr\"odinger equation 
\be
\left[2m_Q -{1\over m_Q}\nabla^2 + V(r,T)\right] \Phi_i(r,T) = M_i(T) 
\Phi_i(r,T).
\label{b1}
\ee
Its solution gives the resulting quarkonium masses $M_i(T)$, with $i=0$ for 
the ground state and $i=1,2,...$ for the subsequent excited states. Here 
$m_Q$ denotes the $c$ or $b$ quark mass, while $V(r,T)$ describes the in-medium
binding potential. To obtain a feeling for the resulting behavior, it is 
helpful to consider the semi-classical limit of eq.\ (\ref{b1}) \cite{r4},
\be
\left[2m_Q +{c\over m_Q r^2} + V(r,T)\right] = E(r,T), 
\label{b2}
\ee
with $c$ a parameter of order unity, to be determined such as to give
correct quarkonium masses. In vacuum, for $T=0$, the binding potential 
is generally assumed to have the Cornell form
\be
V(r,T=0) = \sigma r - {\alpha\over r},
\label{b2a}
\ee
where $\sigma$ denotes the string tension and $\alpha$ the Coulombic
running coupling. For illustration, we concentrate for the moment
on the strong coupling form and neglect the Coulombic term; later on, we
shall include it. 

\medskip

\begin{figure}[htb]
\centerline{\epsfig{file=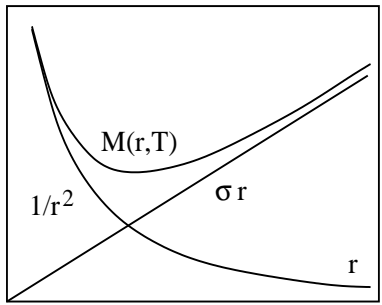,width=5cm,height=4cm}}
\caption{Semi-classical form of quarkonium binding in vacuum}
\label{Fb1}
\end{figure}

In vacuum, we thus set $V(r,T=0)=\sigma r$ and minimize the energy
$E(r,T=0)$ with respect to $r$. This yields
\be
r_0 = \left({2c \over m_Q \sigma}\right)^{1/3}
\label{b3}
\ee
for the vacuum ground state $\Q$ separation and
\be
M_0(\Q) =  2m_Q + {3\over 2} \left({2c\sigma^2 \over m_Q}\right)^{1/3}
\label{b4}
\ee
for the corresponding
vacuum ground state mass. The behavior of $[E(r,0)-2m_Q]$ is
illustrated in Fig.\ \ref{Fb1}; the minimum arises from the competition
between the kinetic energy term $c/mr^2$ and the binding potential
$\sigma r$. With $\sigma=0.2$ GeV$^2$, $m_c=1.3$ GeV and $c=1.3$, this 
yields $r_0(\C) = 0.43$ fm for the diameter and $M_0(\C)= 3.2$ Gev for the 
mass of the charmonium ground state. Similarly, the bottomonium sector gives 
with $m_b= 4.6$ GeV and $c=1.1$ the values  $r_0(\B) = 0.27$ fm and
$M(\B) = 9.6$ GeV. The bound states thus arise for parameter values for 
which the competition comes to a draw.

\medskip

\medskip

We now place the given $\Q$ pair into a deconfined medium of temperature $T$. 
This will have two distinct effects: the constituents of the medium will 
modify the interaction between the static quarks, and they will interact 
with each of the static quarks individually. The binding of the pair is
mediated by an exchange of virtual gluons, and this exchange is modified by 
their interaction with the on-shell constituents of the medium. As a result, 
the binding force itself is modified. On the other hand, the on-shell
constituents of the medium can also interact directly and individually 
with the $Q$ and the $\bar Q$. For sufficiently large separation distance, 
this leads to the formation of polarization clouds around each heavy quark.
On one hand, this requires an energy input to provide the increase of their
effective mass; on the other hand, the entropy of the overall system grows. 
This entropy increases over a range of $r$, and the resulting entropic force 
acts to pull $Q$ and $\bar Q$ apart.
Let us look at these possible modifications, starting with a static $\Q$ 
pair: in other words, we first consider the effect on the potential term only. 

\medskip

\begin{figure}[htb]
\centerline{\epsfig{file=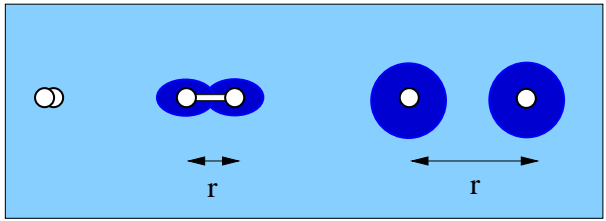,width=8cm}}
\caption{$\Q$ pair separation and polarization cloud formation}
\label{Fb2}
\end{figure}

In the limit of small $r$, the two heavy quarks form a color neutral entity, 
so the medium does not see them, nor do they see the medium. The only 
energy difference between our medium and one without a $\Q$ pair is thus 
the mass of the small $\Q$ system. To further separate $Q$ and $\bar Q$, 
work has to be done against the interquark binding force. Increasing 
the temperature of the medium will modify the binding, and this of course 
has to be taken into account. Once the quarks are sufficiently separated, 
their color charges begin to induce polarization effects in the medium, 
first forming a cloud around the pair. With further separation, the quarks 
interact less strongly with each other, but the medium now forms clouds 
around the individual $Q$ and $\bar Q$. The overall picture is schematically 
illustrated in Fig.\ \ref{Fb2}.

\medskip

\medskip

For a small imagined volume containing the $\Q$ pair, we thus note two
consequences. For a separation increase $\Delta r$, the string energy
grows by $\sigma \Delta r$; in addition, the effective mass of each heavy 
quark begins to increase, due to the onset of polarization. This 
means that the overall energy in the volume increases. On the other hand, 
the rearrangement of the constituents of the medium to form polarization
clouds around each heavy quark results as well in an increase of the relevant 
entropy. In view of this interplay of energetic and entropic effects, the 
form of the potential to be used in the Schr\"odinger equation (\ref{b1}) 
has been the subject of considerable discussion. Proposals range from the 
free energy, $V(T,r) = F(T,r)$, to the internal energy, $V(T,r)=U(T,r)$, 
and include even arbitrary combinations of the two \cite{p1,p3,p4,p4a,
p5,p5a,r4a}.
We want to show here that a careful study of the underlying forces can 
resolve this issue to some extent.
  
\medskip

In the next section, we shall first address the behavior of the system in
the strong coupling regime, where the thermodynamics can be formulated in
terms of a remnant string tension as binding. Following that, we turn to
the weak coupling regime, in which the interquark forces are of Coulombic
nature. In a final section, we shall then address the specific effects
arising in the region dominated by critical behavior.

{\large \section{The Strong Coupling Regime}}
 
To study this region in more detail, we consider a specific simple
model, based on a remnant string tension $\sigma$ as binding force in a
deconfined plasma. To begin, we assume the medium to give rise to 
a screening mass $\mu(T) = {\rm const.} 
~\!T$; the modification of $\mu(T)$ by critical behavior near the 
deconfinement point $T_c$ will be addressed later on. For the free energy 
of the pair in the medium we assume the form \cite{r4}
\be
F(r,T) = \sigma r \left[ {1 - e^{-\mu r} \over \mu r} \right]
= {\sigma \over \mu} [1 - e^{-x}],~x=\mu r,
\label{b5}
\ee   
with a screening factor $(1 - e^{-\mu r})/\mu r$ based on one-dimensional
QED \cite{Dixit,r5}. We have here neglected short distance
Coulombic effects on the quark binding; we shall also return to these later.
The expression (\ref{b5}) specifies the free energy difference between a 
plasma containing a $\Q$ pair and one at the same temperature without such a 
pair. All further thermodynamic quantities will also describe the 
corresponding differences; we will not explicitly note that in each case 
in the following. 

\medskip

The corresponding entropy becomes
\be
T S(T,r) = - T\left({\partial F \over \partial T}\right)_r 
= \left({\sigma \over \mu}\right)  \left[1 - (1+x) e^{-x} \right];
\label{b6}
\ee
with the assumed screening mass linear in $T$, so that $T (d\mu/dT)=\mu$.
We note that in the short distance limit, the entropy vanishes, as
expected. In the large distance limit at constant temperature, on the other 
hand, it attains a constant value, corresponding to the entropy of the
two polarization clouds.

\medskip

\begin{figure}[htb]
\centerline{\epsfig{file=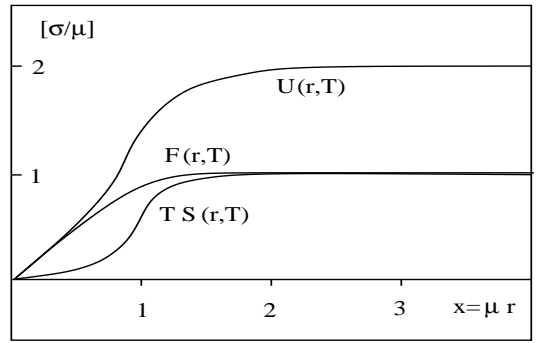,width=7cm,height=4.5cm}}
\caption{Strong coupling form of thermodynamic $\Q$ potentials}
\label{Fb3}
\end{figure}

From the free energy and the entropy, we obtain the total (internal) energy 
difference for the system
\be
U(T,r) = F(T,r) + T S(T,r)= \left({\sigma \over \mu}\right) 
\left[2 - (2+x) e^{-x} \right],
\label{b7}
\ee
consisting of one term accounting for the work done against the string
tension and one for the formation of the polarization clouds. In the
present model, each of the terms contributes in the large distance limit
an equal amount $\sigma/\mu$ to the total energy. In Fig. \ref{Fb3} the 
resulting behavior of free energy, entropy and total energy are illustrated.
The total energy initially increases because the free energy does, but for
$x \gsim 1$, it does so because of the formation of the polarization masses.

\medskip

\medskip

The resulting pressure $P(T,r)$ is given by
\be
P(T,r) = - \left({\partial F \over \partial r}\right)_T = -\sigma e^{-x}
\label{b8}
\ee
It is negative, indicating that the $\Q$ binding is attractive. Its
absolute value decreases with increasing $\mu$, indicating that growing 
temperature enhances screening and hence weakens the binding, causing it 
to vanish in the high temperature limit. Similarly, at fixed $\mu$, color 
screening strongly reduces the binding for separation distances $r > r_D 
= 1/\mu$, when quark and antiquark can no longer communicate. 

\medskip

\begin{figure}[htb]
\centerline{\epsfig{file=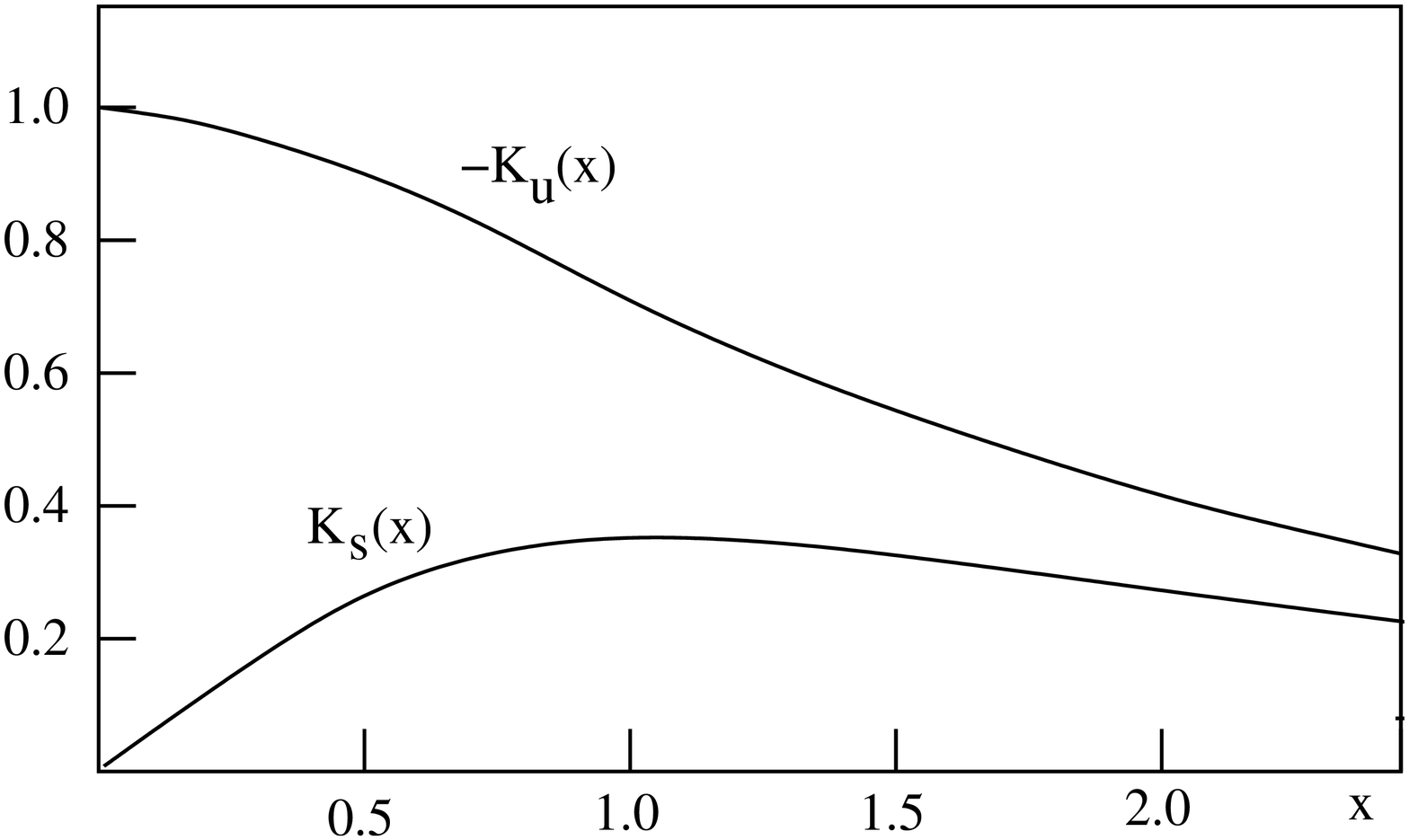,width=7cm,height=5cm}
\hskip1cm \epsfig{file=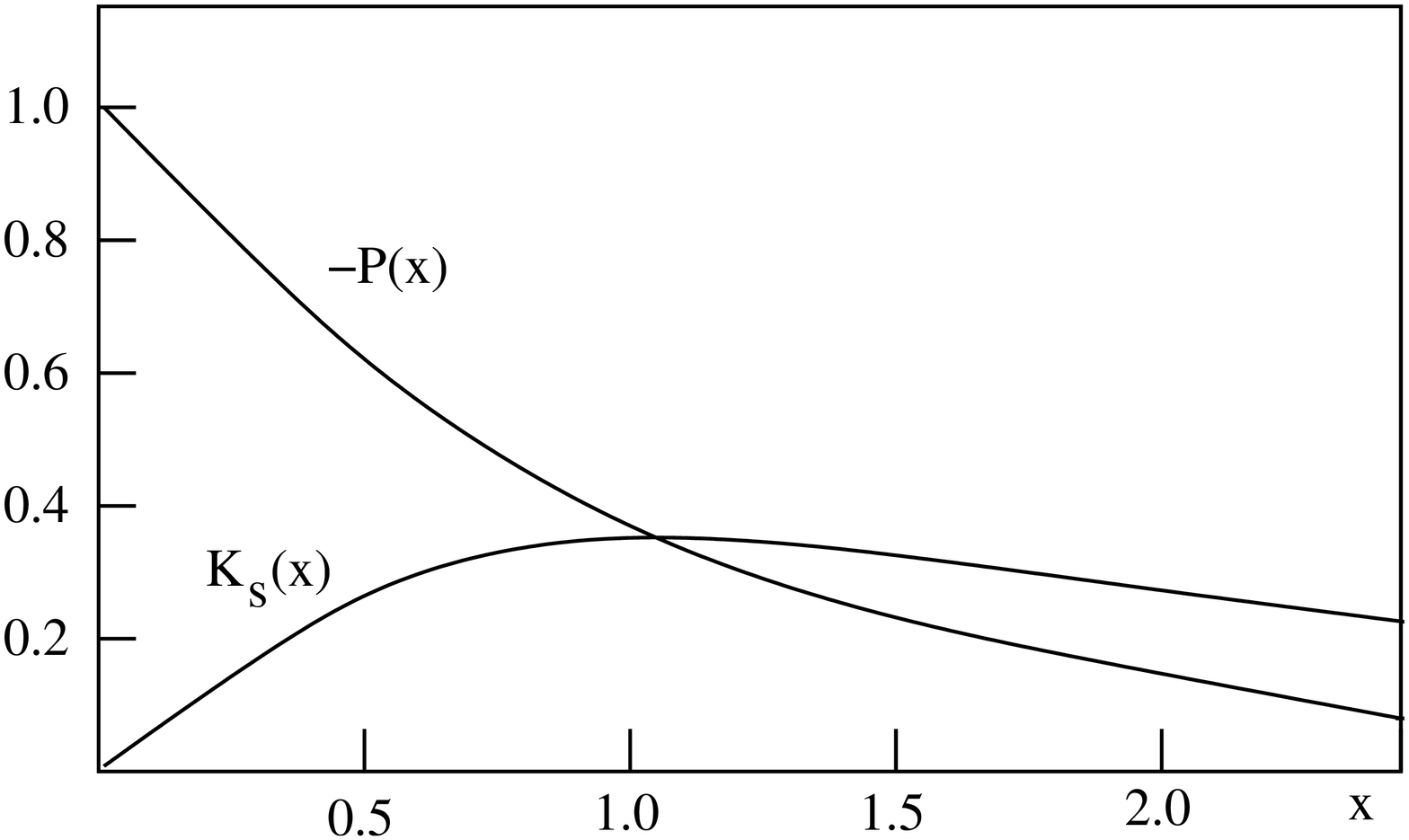,width=7cm,height=5cm}}
\caption{Strong coupling forms 
of energetic and entropic forces and of pressure,
in units of $\sigma$}
\label{strongforce}
\end{figure}

The pressure specifies the force acting on the $\Q$ pair. It consists of
two distinct terms; from the relation $F=U-TS$ between free energy, internal
energy and entropy we have
\be
P(T,r) = 
- \left({\partial U \over \partial r}\right)_T
+T \left({\partial S \over \partial r}\right)_T
= K_u(T,r) + K_s(T,r).
\label{b9}
\ee
Here $K_u=-(\partial U/\partial r)_T$ denotes the {\sl energetic force} and
$K_s=(\partial S/\partial r)_T$ the {\sl entropic force}. Both 
$(\partial U/\partial r)_T$ and $(\partial S/\partial r)_T$ increase
for increasing $r$. The former increases first by the energy spent to separate
the pair against the attractive string tension and then by the energy needed
to form the polarization clouds acquired by the separated quarks. The 
latter does so because the increase in the constituent density of the 
medium around $Q$ and $\bar Q$ increases the overall entropy. The two 
forces $K_u$ and $K_s$ thus act in opposite directions, with $K_u$ attractive 
and $K_s$ repulsive; hence they largely cancel each other. 
For the above model, we obtain
\be
P(T,r) = \left[- \sigma (1+x)e^{-x}\right] + \left[\sigma x e^{-x}\right]
= -\sigma e^{-x},
\label{b10}
\ee
with the two terms in square brackets denoting energetic and entropic force,
respectively. Their behavior as function of $r$ at fixed temperature is 
illustrated in Fig.\ \ref{strongforce}. At short distances, the energetic
force dominates, while in the large distance limit, the ratio of the two
forces approaches unity. 

\medskip

The overall force acting on the $\Q$ pair is given by the pressure;
since it is obtained from the free energy, $F(T,r)$ is the relevant binding
potential. The steeper potential increase and hence stronger binding obtained 
when $U(T,r)$ is used in the Schr\"odinger equation \cite{p1,p3,p4,p5}
arise only because the repulsive entropic force is ignored; it in fact
compensates the stronger attraction suggested by $U$. 
The overall effect on the $\Q$ is due to a combination of the two opposing 
forces, and this leads to $F(T,r)$. Nevertheless, one may still question the
validity of the two-body Schr\"odinger approach in regions in which the
repulsive entropic force is significantly larger than overall pressure, i.e.,
for $x \gsim 1$. In this region, the combined strength of the interaction
between the $Q$ and the on-shell constituents of the medium is stronger than 
that with the $\bar Q$.

\medskip

Before turning to the weak coupling regime, we note that through the 
Maxwell relation
\be
T\left({\partial P \over \partial T}\right)_r= 
T\left({\partial S \over \partial r}\right)_T= 
K_s(T,r) = \sigma x e^{-x} 
\label{s6}
\ee
the temperature dependence of the pressure at fixed $r$ is in fact
determined by the entropic force alone. 

{\large \section{The Weak Coupling Regime}}

Here the interquark potential is of screened $1/r$ type, so that the difference
in free energy between a system with and one without a $\Q$ pair is given by
\be
F(T,r) = - \alpha {e^{-\mu r} \over r} - \alpha \mu;
\label{w1}
\ee 
again $\mu(T) \sim T$ is the color screening length in the medium,
$r_D=1/\mu$ the Debye radius. The second term above accounts for the
effect of the polarization clouds at infinite separation, and $\alpha$
denotes the (running) Coulombic coupling. Defining $x = \mu r$, we rewrite 
eq.\ (\ref{w1}) as
\be
F(T,r) = - \alpha \mu \left[ 1 +{ e^{-x} \over x}\right].
\label{w1a}
\ee 

\medskip

\begin{figure}[htb]
\centerline{\epsfig{file=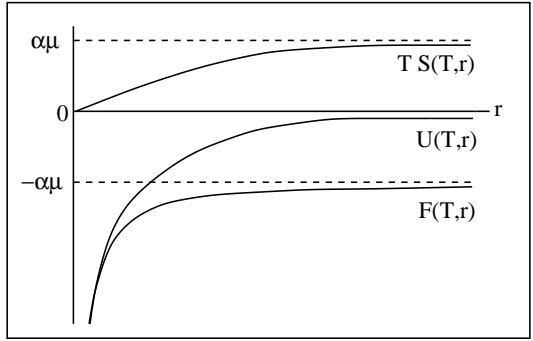,width=7cm}}
\caption{Weak coupling form of thermodynamic $\Q$ potentials}
\label{pot}
\end{figure}

The corresponding entropy difference is given by
 \be
T S(T,r) = - T\left({\partial F \over \partial T}\right)_r 
= \alpha \mu \left[1 - e^{-x} \right];
\label{w2}
\ee
it vanishes for $r=0$ and also approaches the individual cloud contributions 
in the large distance limit. From free energy and entropy we obtain the total
energy difference
\be
U(T,r) = F(T,r) + T S(T,r) = -\alpha \mu \left[1 + {1 \over x}\right]
e^{-x}.
\label{w3}
\ee
It is seen to approach the free energy form for small $r$, while vanishing
in the large distance limit, where free energy and entropy cancel each
other: the work done in separating $Q$ and $\bar Q$ is balanced by the 
increase of entropy. The behavior of the three thermodynamic potentials is 
illustrated
in Fig.\ \ref{pot}. 

\medskip

\medskip

From the free energy, we obtain for the resulting pressure 
\be
P(T,r) = - \left({\partial F \over \partial r}\right)_T = 
- \alpha \mu \left[{1 \over \mu r} + 1 \right]{e^{-\mu r} \over r}=
- \alpha \mu^2 \left[{1 \over x^2} + {1 \over x} \right]e^{-x}.
\label{w4}
\ee
It is negative, indicating that also here there
is a force acting to contract the
pair. Its absolute value decreases with increasing separation distance $r$
and vanishes in the large distance limit.

\medskip

The pressure again consists of an entropic and an energetic force. The former 
is given by
\be
K_s(T,r) = T \left({\partial S \over \partial r}\right)_T 
= \alpha \mu^2 e^{-x}.
\label{w5}
\ee
It is positive, enhancing the dissociation of the pair. The corresponding
energetic force becomes
\be
K_u(T,r) = - \left({\partial U \over \partial r}\right)_T 
= - \alpha \mu^2 \left[{1 \over x^2} + {1 \over x}+1 \right]e^{-x}.
\label{w6}
\ee
It dominates in the short distance limit, while for large distances
entropic and energetic forces become equal and cancel each other, resulting
in a vanishing overall pressure. --
Through the Maxwell relation
\be
T \left({\partial P \over \partial T}\right)_r=
T \left({\partial S \over \partial r}\right)_T=K_s(T,r) 
= \alpha \mu^2 e^{-x}
\label{w7}
\ee
the entropic force again determines the temperature variation of the 
pressure. 

\bigskip

{\large \section{The Critical Regime}}

\begin{figure}[htb]
\centerline{\epsfig{file=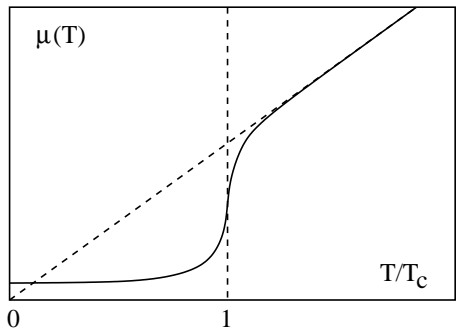,width=6cm}}
\caption{The temperature variation of the screening mass $\mu(T)$}
\label{screen}
\end{figure}

So far, we had assumed the screening mass to increase linearly in $T$.
This assumption evidently becomes incorrect when the temperature drops to
the region just above the deconfinement point. The behavior of $\mu(T)$
is illustrated schematically in Fig.\ \ref{screen}; approaching $T_c$,
$\mu(T)$ abruptly drops to a small value determined by string breaking
in the confined phase. This pattern affects in particular the thermodynamic
potentials involving $d\mu/dT$, which increases very sharply near $T_c$ and 
in principle could even diverge there.   

\medskip

For $\mu(T)$ decreasing to very small values, the free energy 
\be
F(r,T) = {\sigma \over \mu} [1 - e^{-x}]- \alpha \mu 
\left[ 1 +{ e^{-x} \over x}\right]
\label{c1}
\ee   
essentially falls back to the Cornell potential
\be
F(T,r) \simeq \sigma r - {\alpha \over r}.
\label{c2}
\ee
Here and in the following, we shall consider the full thermodynamic 
potentials, combining the expression for the strong and weak coupling limits.
Corresponding to eq.\ (\ref{c1}), the pressure becomes
\be
P(T,r) = -\left[\sigma + \alpha \mu^2\left({1\over x} + {1 \over x^2}
\right)\right]e^{-x}
\label{c3}
\ee 
In the small distance limit, it also reduces to the attractive force in
vacuum,
\be
P(T,r) \simeq -\left(\sigma + {\alpha \over r^2}\right).
\label{crit-pres}
\ee
In both cases, medium effects become significant only at large $r$,
where the screening factor $\exp\{-\mu r\}$ becomes significantly less
than unity and begins to play a role. 

\medskip

In contrast to free energy and pressure, the entropy and the corresponding
force contain temperature derivatives
of the screening mass, $d\mu/dT$, which near the critical deconfinement
temperature increase very sharply. The entropy is given by
\be
S(T,r) = (d\mu/dT)\left({\sigma \over \mu^2}\left[1-(1+x)e^{-x}\right]
+ \alpha\left[1-e^{-x}\right]\right), 
\label{full-en}
\ee
and the entropic force becomes
\be
K_s(T,r) = T(d\mu /dT) \left({\sigma \over \mu} x e^{-x} + \alpha \mu e^{-x}
\right).
\label{full-force}
\ee 
For completeness, we note that
\be
U(T,r) = T(d\mu / dT) 
\left({\sigma \over \mu^2}\right)\left\{
\left[2 - (2+x)e^{-x}\right] 
- {\alpha\over \mu} \left[1 + {1\over x}\right]e^{-x}\right\}
\label{full-int}
\ee
gives the total internal energy of the system. It again contains the 
temperature derivative of the screening mass and thus strongly reflects 
the critical behavior near $T_c$. 

\medskip

\begin{figure}[htb]
\centerline{\epsfig{file=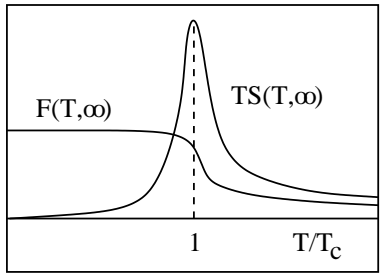,width=5cm}}
\caption{Large distance limits of free energy and entropy vs.\ temperature}
\label{comp1}
\end{figure}

To obtain a first idea of the role of critical behavior, we compare in
Fig.\ \ref{comp1} the large distance forms of free energy and entropy.
It is evident that near the critical point, there is a sudden increase of
the entropy, caused by the increase or divergence of the correlation length,
which now connects ever larger regions of the medium.

\medskip

\begin{figure}[h]
\centerline{\epsfig{file=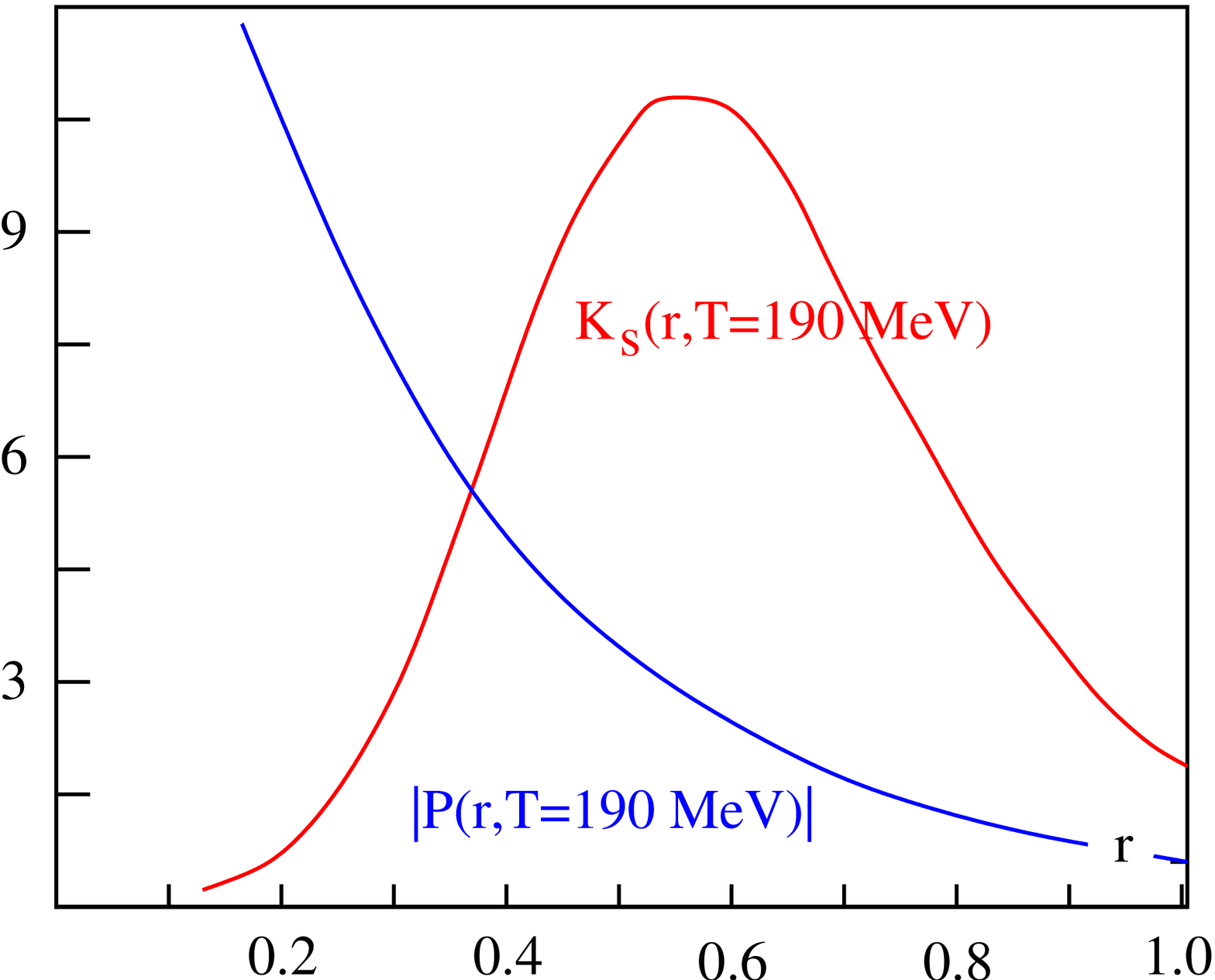,width=5.7cm}\hskip1cm
\epsfig{file=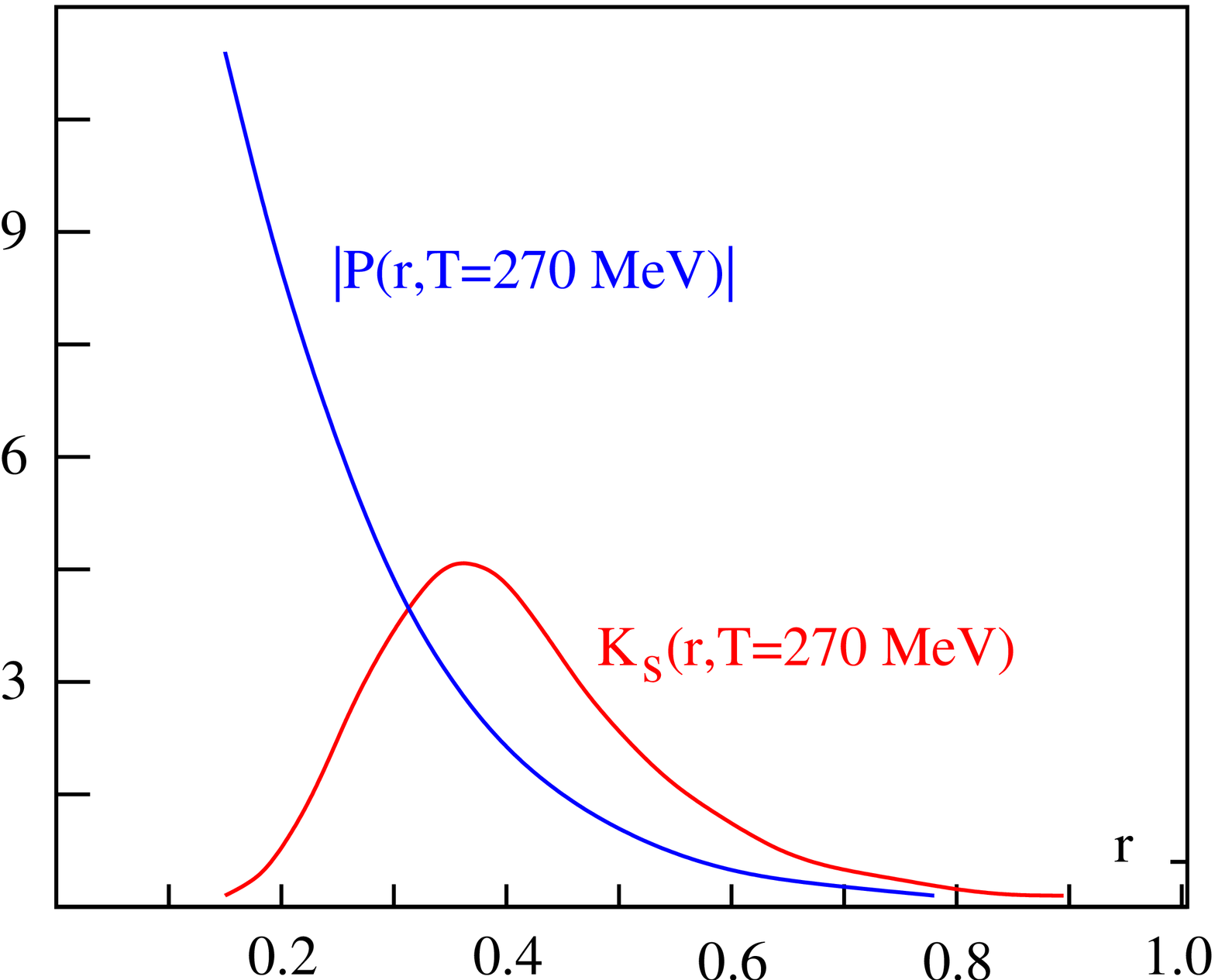,width=5.7cm}}
\caption{The pressure $P(r,T)$ (blue) and the entropic force $K_s(r,T)$
(red) as function of $r$ for two temperature values \cite{r7,r8,r9,r10a}; 
the forces are given in units of [fm$^{-2}$].}
\label{lat}
\end{figure}

To study the actual behavior of the relevant forces, we turn to finite 
temperature lattice studies \cite{r7,r8,r9}. In Fig.\ \ref{lat} we show
results for two different temperatures, $T=190$ MeV and $T=270$ MeV. The
lower temperature is quite close to the deconfinement value of about
160 MeV. It is seen that here the entropic force becomes strongly enhanced 
relative to the pressure, and in range of distances around 0.5 fm it is 
in fact much larger. The low pressure values thus reflect the cancellation
of entropic and energetic forces in this region.

\medskip

The striking and perhaps surprising conclusion thus is that the divergent
critical effects near $T_c$ in fact cancel out for the $\Q$ binding; the
overall potential between $Q$ and $\bar Q$ only notices a sudden decrease
at the critical point. Here the (negative) pressure is simply reduced 
considerably, allowing an easier dissociation than in vacuum. In a potential
approach based on equilibrium thermodynamics and the two-body Schr\"odinger 
equation (\ref{b1}), there is thus no special entropic critical effect 
causing stronger quarkonium dissociation.

\medskip

Nevertheless, it is evident from Figs.\ \ref{comp1} and \ref{lat} that both 
entropy and entropic force show sharp increases near $T_c$. This indicates
that in the relevant $r$ regions, the interaction between a heavy quark and
the combined constituents of the medium becomes much stronger than that
between $Q$ and $\bar Q$. This may well be a signal that here a two-body
potential approach is no longer applicable, allowing an enhanced collective
entropic quarkonium dissociation \cite{r2}.

{\large \section{Concluding Remarks}}

In summary, we conclude that {\sl if} quarkonium binding and dissociation is
to be studied in terms of a potential approach, the relevant potential is the
free energy difference $F(T,r)$ between a medium with and one without a 
$\Q$ pair. The stronger increase of the internal energy $U(T,r)$ with 
increasing separation distance $r$ is compensated by equally strong
contributions from the repulsive entropic force, opposing rather than 
enhancing 
any binding. Hence $U$ as a potential only makes sense when these are not 
present, i.e., when $U \simeq  F$.

\medskip

The conclusion that the free energy is the appropriate potential for $Q$ 
binding is also in accord with previous work comparing real and imaginary 
time $\Q$ correlators in a thermal medium \cite{weak}. Very recently, it was
obtained as well in a potential derivation from real-time Wilson loop studies 
\cite{r10}.

\medskip

At first sight, the long-standing question, $F$ or $U$?, thus appears to be 
resolved in favor of $F$. However, as long as the role of larger distance 
repulsive 
entropic effects is not fully clarified, it remains conceivable that a 
potential approach as such only makes sense in that $(T,r)$ region where 
$F \simeq U$. Finally, we note that we have here addressed only the real
part of the potential. In more formal studies \cite{Brambilla}, the important 
issue of its imaginary part, reflecting Landau damping and singlet-octet 
transitions, can be addressed as well. This aspect is beyond the scope of the
present work; for recent studies and further literature, see 
\cite{weak,r10,Brambilla,r11,r12}.

\bigskip

{\large \centerline{\bf Acknowledgements}}

\bigskip

The paper of D.\ Kharzeev \cite{r2} was a crucial stimulus for this work,
even though its topical focus differs from ours. It is a pleasure to thank 
him for sending it to me prior to publication and for several stimulating 
and helpful exchanges. I also want to thank F.\ Karsch for critical readings 
of earlier drafts of this work and providing most helpful input. Further
thanks go to O.\ Kaczmarek and P. Petreczky for stimulating discussions,
and to N.\ Brambilla and A.\ Vairo for helpful comments. 
The hospitality of the Institute for Nuclear Theory, University of Washington, 
and the partial support of the Department of Energy during the start of this 
work are gratefully acknowledged.


\bigskip


\begin{thebibliography}{99}

\bibitem{r1a} R.\ M.\ Neumann, Am.\ J.\ Phys.\ 48 (1980) 48.
\bibitem{r1b} E.\ P.\ Verlinde, JHEP 1104 (2011) 029.
\bibitem{r1c} P.\ G.\ O.\ Freund, arXiv:1008.4147.
\bibitem{r2} D.\ E.\ Kharzeev, arXiv:1409.2496.
\bibitem{r3} T.\ Matsui and H.\ Satz, \PL B 178 (1986) 416.
\bibitem{r4} F.\ Karsch, M.-T.\ Mehr and H.\ Satz, \ZP C 37 (1988) 617.
\bibitem{p1} S.\ Digal, P.\ Petreczky and H.\ Satz, \PL B 514 (2001) 57.
\bibitem{p3} E.\ Shuryak and I.\ Zahed, \PR D {\bf 70}, 054507 (2004).
\bibitem{p4} C.-Y.\ Wong, \PR C 72 (2004) 034906. 
\bibitem{p4a} C.-Y.\ Wong, \PR C 76 (2007) 014902.
\bibitem{p5} W.\ Alberico et al., \PR D 72 (2005) 114011.
\bibitem{p5a} W.\ Alberico et al., \PR D 75 (2007) 074009.
\bibitem{r4a} H.\ Satz, {\sl Extreme States of Matter in Strong Interaction
Physics}, Lecture Notes in Physics 841, Springer Verlag, 2012; chapter 8.
\bibitem{Dixit} V.\ V.\ Dixit, Mod.\ Phys.\ Lett.\ A 5 (1990) 227.
\bibitem{r5} S.\ Digal, P.\ Petreczky and H.\ Satz, \EPJ C 43 (2005) 71. 
\bibitem{r7} O. Kaczmarek and F. Zantow, PR D 71 (2005) 114510
\bibitem{r8} O. Kaczmarek and F. Zantow, PoS LAT 2005:177 2006.
\bibitem{r9} O.\ Kaczmarek and F.\ Zantow, arXiv:0710.0498.
\bibitem{r10a} H.\ Satz, J.\ Phys. G36 (2009) 064011.
\bibitem{weak} A.\ Beraudo, J.-P.\ Blaizot and C.\ Ratti,\NP A806 (2008) 312.
\bibitem{r10} Y.\ Burnier, O.\ Kaczmarek and A.\ Rothkopf, arXiv:01410.7311
  and 01411.3141.
\bibitem{Brambilla} N.\ Brambilla et al., arXiv:0804.0993.
\bibitem{r11} M.\ Laine et al, JHEP 0703 (2007) 054.
\bibitem{r12} M.\ Laine, O.\ Philipsen and M.\ Tassler, JHEP 0709 (2007) 066.



\end{thebibliography}
\end{document}